# A Hybrid Finite Element and Material Point Method for Modeling Liquefaction-Induced Tailings Dam Failures

Brent Sordo, Ph.D.[1], Ellen Rathje, Ph.D.[2], Krishna Kumar, Ph.D.[3]

[1]Former Ph.D. Candidate, Department of Civil, Architectural, and Environmental Engineering, The University of Texas at Austin, 301 E Dean Keeton St, Austin, TX 78712; E-mail: bsordo@utexas.edu
[2]Professor, Department of Civil, Architectural, and Environmental Engineering, The University of Texas at Austin, 301 E Dean Keeton St, Austin, TX 78712; E-mail: e.rathje@mail.utexas.edu
[3]Assistant Professor, Department of Civil, Architectural, and Environmental Engineering, The University of Texas at Austin, 301 E Dean Keeton St, Austin, TX 78712; E-mail: krishnak@utexas.edu

**ABSTRACT**

This paper presents a hybrid Finite Element Method (FEM) and Material Point Method (MPM) approach for modeling liquefaction-induced tailings dam failures from initiation through runout. We apply this method to simulate the 1978 Mochikoshi tailings dam failure, which occurred due to seismic loading and liquefaction during an earthquake. Our approach leverages FEM to capture the initial failure mechanism and MPM to simulate the subsequent runout, exploiting the strength of each method in their respective phases of the failure process. We investigate the impact of the FEM-to-MPM transfer time on runout results, identifying an optimal transfer window. This window begins when liquefaction reaches a critical depth to fully trigger the failure and ends before excessive mesh deformation occurs. Our findings demonstrate that the properties of the liquefied tailings significantly influence runout predictions. Notably, we achieve runout distances comparable to the case history only when incorporating additional strain-softening beyond the initial liquefaction-induced strength reduction. Our results demonstrate that the hybrid FEM-MPM method effectively models tailings dam failures associated with complex failure mechanisms and large runouts. This approach offers a promising tool for predicting the runout of seismic liquefaction-induced tailings dam failures, improving risk assessment and mitigation strategies in tailings dam management.

## INTRODUCTION

Tailings dams are among the largest engineered structures on earth and are designed to retain massive quantities of potentially saturated tailings. Their failures can release massive flow slides, posing a severe threat to local communities and natural environments (Kossoff et al., 2014; Adamo et al., 2020). Unfortunately, tailings dam failures are unacceptably frequent, averaging more than three annual occurrences, even in well-regulated countries (Fourie et al., 2022). In response to past failures, many countries have begun advising or requiring engineering analysis of the stability of the dam, the potential volume released by a breach, and inundation maps indicating areas that would be affected by a dam failure (Martin et al., 2015; Larrauri and Lall, 2018; ICMM, 2020).

We present a novel hybrid Finite Element Method (FEM) and Material Point Method (MPM) approach to model liquefaction-induced tailings dam failures from initiation through runout. Our method uniquely combines the strengths of FEM in modeling complex failure mechanisms with the ability of MPM to simulate large deformations, overcoming the limitations of each method used individually. We demonstrate this approach by successfully modeling the 1978 Mochikoshi tailings dam failure, providing insights into the failure mechanism and runout behavior.

The exact mode and characteristics of the initial dam breach play a key role in determining the extent of the runout (Martin et al., 2015). One potential triggering mechanism for tailings dam failures is earthquake loading, which usually induces failure by liquefying the retained tailings and causing severe strength loss (Lyu et al., 2019). For these types of failures, an accurate seismic response analysis, particularly in predicting the extent of liquefaction, is a critical component of the engineering analysis of a tailings dam.

The Finite Element Method (FEM; Zienkiewicz et al., 2005) has become a popular tool for predicting tailings dam failures, among other phenomena (Griffiths and Taylor, 1999; Geppetti et al., 2022). The FEM is favored over traditional Limit Equilibrium Methods (LEMs) because it accounts for strain compatibility and stress equilibrium, and it models the material behavior with a comprehensive constitutive law. With appropriate constitutive models, the FEM can predict the accumulation of excess pore water pressure during earthquake shaking and the potential for, and extent of, soil liquefaction. Thus, the FEM can identify the failure initiation mechanism for a tailings dam and the initial evolution of its failure surface. While adept at analyzing the initiation of dam failures, the FEM loses accuracy when predicting large deformations due to mesh distortion and entanglement. While re-meshing (Zhang et al., 2016) and displacement field refinement (Zeng and Liu, 2016) mitigate these issues, they require substantial computational effort. Therefore, the inherent limitation of mesh distortion in the FEM precludes it from effectively predicting the runout of flow slides that result from tailings dam failures (Soga et al., 2015; Liang and Zhao, 2018).

The Material Point Method (MPM; Sulsky et al., 1994; Bardenhagen et al., 2000), a hybrid Eulerian-Lagrangian approach, is an alternative numerical approach suitable for modeling large deformation problems. MPM can account for large displacements, such as landslide runouts, without suffering from mesh distortions (Cuomo et al., 2021). However, the integration of stresses via moving material points is less accurate than Gauss integration in the FEM, resulting in severe stress checkerboarding in MPM models (Wang et al., 2021). However, methods have been developed to help reduce these stress oscillations, such as the Generalised Interpolation Material Points (GIMP) method (Bardenhagen and Kober, 2004), the B-spline Material Point Method (Steffen et al., 2008), the Convected Particle Domain Interpolation (CPDI) method (Sadeghirad et al., 2011), and the Composite Material Point Method (CMPM; Acosta et al., 2020). MPM is also limited in its ability to include absorbing boundary conditions. Dashpots (Shan et al., 2021) and dynamic boundary conditions have been utilized in MPM to model seismic response (Kohler et al., 2021; Alsardi and Yerro, 2023; Kurima et al., 2024), but the potential for material points to move away from the boundary cells limits its efficiency for this application. Recent advancements in MPM and other particle methods, such as the development of novel constitutive models (e.g., Wang et al., 2023; Hoang et al., 2024a) and improved boundary condition treatments (e.g., Chen et al., 2022; Hoang et al., 2024b), have addressed some of these limitations. However, challenges remain in accurately capturing the complex mechanisms of liquefaction-induced failures, even with these improvements, and further advancement is necessary to match the full capabilities of FEM (Lu et al., 2023; Hoang et al., 2024b). Thus, while MPM can model large deformations, it has shortcomings in capturing failure initiation due to these issues associated with stresses and dynamic boundary conditions.

While both FEM and MPM have made significant strides in modeling tailings dam failures, neither method alone can effectively capture the entire failure process from initiation through large-deformation runout. To overcome the shortcomings of both approaches, we utilize a hybrid FEM-MPM method (Sordo et al., 2024a) to simulate the entire failure process of earth structures from initiation to runout. In this method, we model the failure initiation via the FEM, capturing the seismic response, liquefaction, and development of the initial failure surface. Then, we utilize a transfer algorithm at a user-specified time to transfer the entirety of the state of the model from FEM to MPM and continue to simulate the runout via MPM. FEM and MPM have been combined to model rainfall-induced landslides by Lu et al. (2023), where FEM modeled the rainwater infiltration and MPM the runout. Furthermore, similar approach using the Finite Difference Method (FDM) with MPM was described by Talbot et al. (2024) and used to model the liquefaction-induced failure of the Lower San Fernando Dam (Seed et al., 1975).

Our previous work (Sordo et al., 2024a) employed the hybrid FEM-MPM method to model gravity-driven granular column collapses and simple slope failures. In this work, we demonstrate the full capability of the method to model slope and dam failures with complex failure mechanisms (e.g., seismic liquefaction). Accordingly, we model the 1978 failure of the Mochikoshi No. 1 Tailings Dam caused by the Izu-Ohshima-Kinkai earthquake in Japan. The tailings dam failed due to the seismic loading and liquefaction of the tailings, releasing a flow slide that extended nearly a kilometer downstream (Ishihara, 1984). Through the analysis of the Mochikoshi tailings dam failure, we demonstrate the development of appropriate constitutive model parameters to capture the failure initiation due to liquefaction, we investigate the influence of the FEM to MPM transition time on the runout results, and we evaluate the effect of strain-softening of the post-liquefaction residual strength of the liquefied tailings.

**HYBRID FEM-MPM PROCEDURE**

The hybrid FEM-MPM procedure consists of the three phases: an initiation phase modeled in FEM, a transfer phase that translates the model geometry, material behavior, and kinematics from FEM to MPM, and a runout phase modeled in MPM (Figure 1). A summary of the FEM-MPM workflow is provided below; additional details can be found in Sordo et al. (2024a).

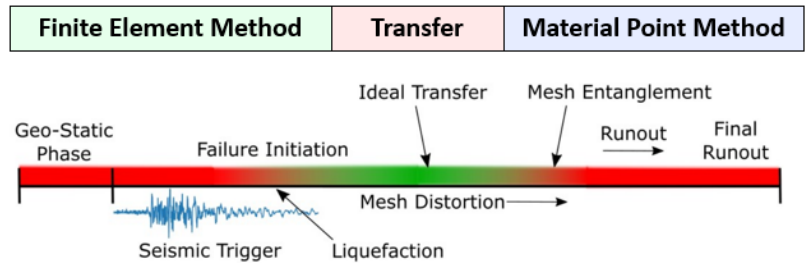

*Figure 1: A schematic timeline of the ideal time to transfer from FEM to MPM.*

The FEM failure initiation phase begins by establishing the initial geostatic stress state. We first compute stresses due to gravity using a linear elastic constitutive model. Subsequently, we transition to a plastic constitutive model to simulate the failure process. To replicate the seismic event, we apply a velocity time history to a Lysmer-Kuhlemeyer dashpot (Lysmer and Kuhlemeyer, 1969) at the base of the model, simulating the dynamic response and triggering the liquefaction-induced failure. The FEM analysis continues until a user-specified transfer time ($t_T$), at which point we switch from FEM to MPM. This transfer time is case-specific and must be chosen before mesh entanglement occurs. As part of this study, we investigate how the choice of transfer time influences the runout results.

The transfer process involves converting the deformed geometries and state variables from FEM into material points for the MPM simulation. We discretize each finite element into material points based on the dimensionality ($D$) of the problem and a user-specified number ($n$) of particles per dimension. For a 2D problem, if $n = 1$ then each element becomes 1 material point, if $n = 2$ each element becomes 4 material points, etc. In general, each $D$-dimensional element is converted into $n^D$ material points. This approach allows for a flexible and systematic conversion of the FEM mesh into MPM particles. In our transfer process, we position the material points at the Gauss locations of the FEM elements, even though the number of Gauss points may differ between the FEM and MPM analyses. At the transfer time ($t_T$), we map the state variables from the FEM phase to the new material point locations. For stresses and strains, which are stored at FEM Gauss points, we use a multiquadratic Radial Basis Function (RBF) with an epsilon value equal to the average distance between Gauss Points for interpolation. For other state variables like displacements and velocities, which are recorded at FEM nodes, we employ the element shape functions to define the values for the transfer. RBF is necessary for the state variables stored at Gauss points because some particles are created beyond the domain encompassed by the Gauss points, so extrapolation is necessary and direct interpolation is insufficient. We allocate a portion of each FEM element's mass and volume to each material point based on the Gauss weight of its location. This approach ensures conservation of mass, volume, kinetic energy, gravitational potential energy, and momentum during the transfer process. For a more detailed explanation of the transfer algorithm, please refer to Sordo et al. (2024a). This algorithm is currently only compatible with quadrilateral elements, but it could be modified to accommodate other element types in the future.

The MPM runout phase involves modeling the large deformation runout resulting from the FEM initialization state. The MPM phase utilizes a computational background grid independent from the FEM mesh, and the material points traverse the background grid. To avoid the need to incorporate absorbing boundary conditions, the input earthquake motion is not included in the MPM phase. Instead, the MPM phase progresses as a gravity driven failure until the runout process is complete.

The analyst must select the transfer time manually based on the condition of the FEM phase (Figure 1). Since the earthquake motion is not included in the MPM phase, the FEM simulation must fully capture the relevant earthquake-induced liquefaction. Transferring too early risks underestimating the runout by underestimating the liquefaction in the model. However, transferring the model too late may result in excessively distorted meshes, and eventually mesh entanglement, as failure develops in the FEM phase, which will then impact the MPM runout (Sordo et al. 2024a). Thus, the ideal transfer time is case dependent and should be selected after the failure mechanism has fully developed but before excessive mesh distortion occurs (Figure 1).

**THE MOCHIKOSHI TAILINGS DAM CASE HISTORY**

*Case History Description*

Construction of the Hozukizawa disposal pond (Figure 2a; Ishihara, 1984) in Mochikoshi, Japan began in 1964 to retain tailings from the Mochikoshi gold and silver mine (Okusa and Anma, 1979; Ishiahra, 1984). The tailings were retained initially by two dams which were founded upon the lightly weathered tuff bedrock; a third dam was added later at a higher base elevation (Okusa and Anma, 1979). As tailings were added to the pond, the dams were raised by the upstream method (Ishihara, 1984), a method of dam raising most susceptible to earthquakes (Adamo et al., 2020). The tailings were composed of sandy silt with a fines content (FC) of ~80% and an average N value of 2 (Ishihara, 1984; Byrne and Seid-Karbasi, 2003). The Izu-Ohshima-Kinkai earthquake (M = 7.0) shook the site on January 14, 1978, causing the failure of dams No. 1 and No. 2 (Ishihara, 1984).

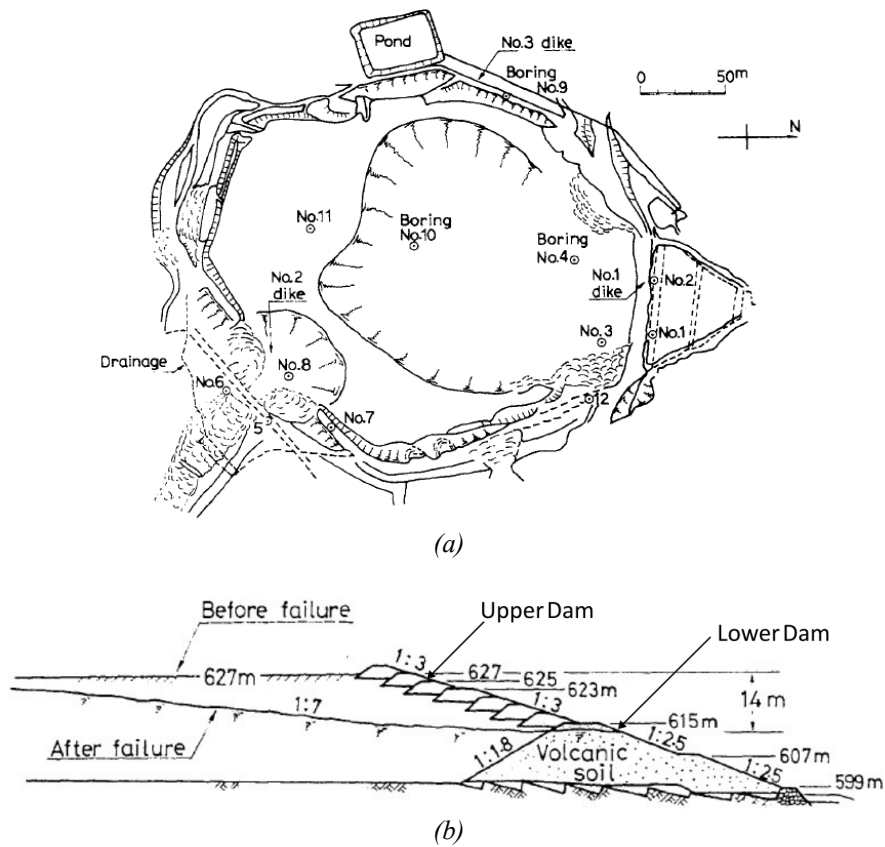

*(a)*

*(b)*

*Figure 2: (a) Plan view and (b) cross-sectional view of Mochikoshi tailing dam No. 1 failure (Reproduced after Ishihara, 1984).*

Within ten seconds of the main shock, a witness described seeing swelling on the face of the No. 1 dam and ultimately a breach of the upper dam (Figure 2b; Ishihara, 1984). Approximately 80,000 m³ of debris flowed through a 73 m wide (Okusa and Anma, 1979) breach in the No. 1 dam and down the mountainside more than 800 m (Ishihara, 1984). The lower portion of the dam, known as the "lower dam", remained in place and retained the remaining tailings. This failure is attributed to the combined effects of seismic loading and strength reduction due to liquefaction in the tailings (Ishihara, 1984). The No. 1 dam is the focus of our study.

Byrne and Seid-Karbasi (2003) used the FDM to model the failure of the No. 1 dam and demonstrate the FDM's capability in accounting for liquefaction. Their analysis utilized the UBCSand constitutive model for the tailings, with the model parameters calibrated to provide liquefaction triggering consistent with predictions from Youd et al. (2001) for an equivalent clean sand normalized SPT blowcount ($N_{1,60-cs}$) of 6. This blowcount was estimated from the post-failure SPT measurements of N ~ 2 and a fines content correction for FC ~ 80%. Without any earthquake recordings available from the vicinity of the dam, they subjected their model to the Caltech B station recording from the 1971 $M_w$ 6.6 San Fernando earthquake. This motion caused significant excess pore pressure generation and liquefaction within the tailings, which resulted in displacements as large as 5 m. However, their mesh became entangled, preventing the simulation of the failure beyond the initial dam breach.

*Limit Equilibrium Analysis of the Mochikoshi Tailings Dam No. 1 Failure*

To evaluate the influence of the extent of liquefaction on the stability of tailings dam No. 1, we create an LEM model of the No. 1 dam. This model is based on the geometry reported by Ishihara (1984; Figure 2b) and is analyzed in Plaxis LE (Bentley, 2021) utilizing the Simplified Bishop Method. The embankment and unliquefied tailings are assigned effective stress Mohr-Coulomb strength properties (Table 1) matching those used by Byrne and Seid-Karbasi (2003). We account for liquefaction by assigning post-liquefaction residual strengths ($s_r$) to liquefied layers based on the in situ initial vertical effective stress ($\sigma'_{vo}$) and $N_{1,60-cs}$ = 6 (Byrne and Seid-Karbasi, 2003). The assigned $s_r$ is key to capturing a tailings dam failure, and field-based empirical relations are commonly used (Sarantonis et al., 2020). We divide the tailings into 5 equal layers ~6 m thick, representative of an increment of $\sigma'_{vo}$ ~ 50 kPa in each layer. A value of $s_r$ representative of the $\sigma'_{vo}$ at the center of each layer is assigned using the empirical relationship of Weber (2015), as tabulated in Table 1. These estimates of $s_r$ were also compared to estimates based on the empirical relationship of Kramer and Wang (2015), and the values were within ± 11%.

Table 1: Material properties for Mochikoshi LEM model.

| Material | Representative $\sigma'_{vo}$ (kPa) | Unit Weight (kN/m³) | φ' (°) | c' (kPa) |
|---|---|---|---|---|
| Embankment | N/A | 16.6 | 35 | 25 |
| Unliquefied Tailings | N/A | 18.3 | 30 | 0 |
| Liquefied Tailings | | | φ (°) | $S_r$ (kPa) |
| Layer 1 | 25 | 18.3 | 0 | 3.0 |
| Layer 2 | 75 | 18.3 | 0 | 5.6 |
| Layer 3 | 125 | 18.3 | 0 | 7.4 |
| Layer 4 | 175 | 18.3 | 0 | 9.1 |
| Layer 5 | 225 | 18.3 | 0 | 10.6 |

To assess the impact of liquefaction depth on dam stability, we perform multiple analyses. We progressively assign residual strengths to additional layers, starting from the uppermost (layer 1) and moving downward to layer 5, simulating increasing depths of liquefaction. With no liquefied layers the factor of safety (FS) is 1.91, and this FS decreases as liquefaction extends deeper (Figure 3a). The FS falls below 1.0 when liquefaction extends to a depth of ~12 m from the surface of the pond (i.e., Layers 1-2 liquefied), and the FS reaches a minimum of 0.54 when the depth of liquefaction extends to ~24 m (i.e., Layers 1-4 liquefied). Liquefaction of the deepest layer has no further effect on the FS. Figure 3b shows the critical failure surface when all the tailings have liquefied. This failure surface extends to a maximum depth of 18.5 m and only passes through the upper dam, which is consistent with the field observations that the lower dam remained in place (Figure 2b).

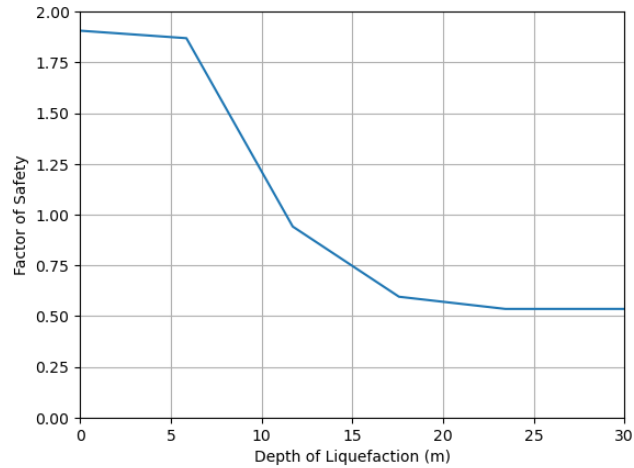

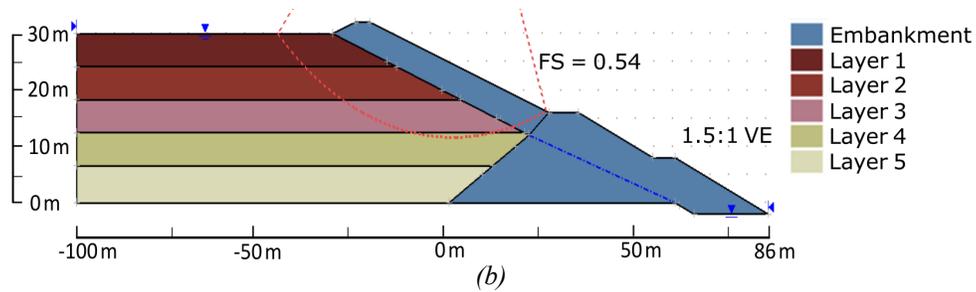

*Figure 3: (a) Factor of safety of the Mochikoshi LEM model versus depth of liquefaction and (b) the failure surface when all tailings are liquefied with minimum FS = 0.54.*

This LEM analysis provides initial insights into the failure mechanism. Liquefaction-induced strength reduction in the tailings leads to a breach in the upper dam, while the lower dam remains stable due to its greater strength and thickness. The analysis suggests that failure (FS = 1.0) may begin when liquefaction reaches a depth of ~12 m. However, the dam becomes increasingly unstable (FS decreases) as liquefaction extends deeper reaching a critical depth of ~18.5 m. Our goal in the FEM phase of the hybrid model is to capture this failure mode.

**FEM PHASE OF HYBRID MODEL TO PREDICT FAILURE INITIATION**

We simulate the initiation phase of our hybrid FEM-MPM model of the Mochikoshi tailings dam using the open-source FEM program OpenSees (McKenna, 1997). The model geometry is discretized into a finite element mesh (Figure 4), using the GiD pre-processing module (Ribó et al., 1999). The model replicates the geometry reported by Ishihara (1984) and rests on a 5-m thick foundation layer of dense sand. A structured mesh is used in the tailings and foundation, but the embankment features an unstructured mesh due to its irregular geometry. Elements near the dam, whether structured or unstructured, are typically 1 m x 1 m, but the elements farther from the dam, near the center of the pond, are widened to 10 m (aspect ratio = 0.1). The model uses SSPquadUP elements (McGann et al., 2012), which are reduced-order, linear, quadrilateral elements with four nodes and a single central Gauss point, and a timestep of 2e-3 s. The left and right boundaries of the model include free-field columns, and a Lysmer-Kuhlemeyer (1969) dashpot is connected to the bottom with properties to model bedrock with $V_s = 760$ m/s. The phreatic surface is located at the surface of the tailings pond, and along the base of the upper dam and extending to the toe within the embankment.

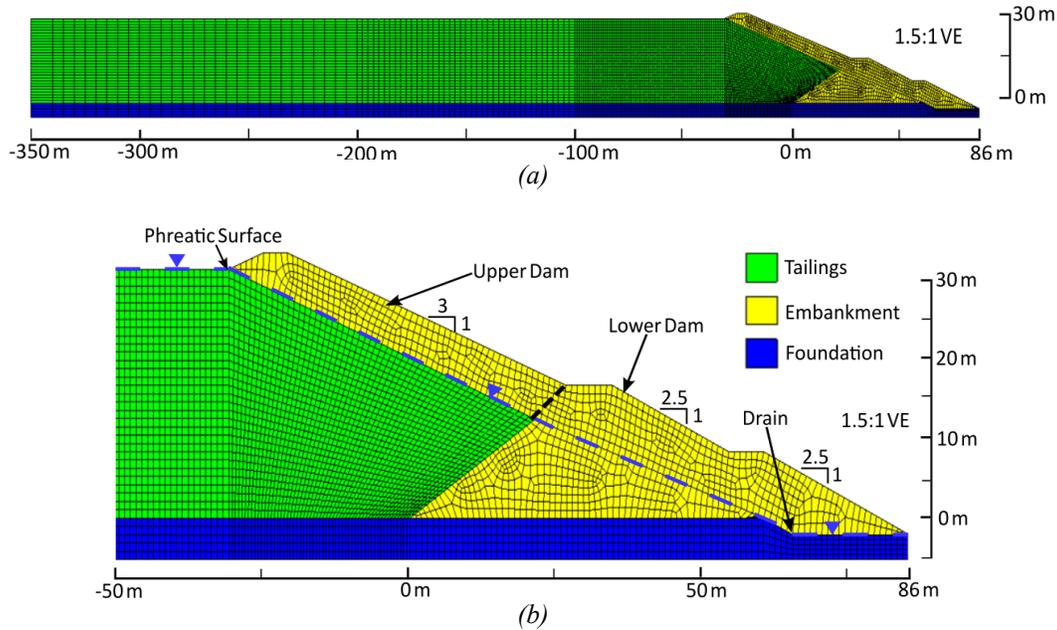

Figure 4: (a) Full view of the mesh of the FEM phase of the Mochikoshi hybrid model and (b) focused view of the unstructured mesh of the dam.

*Constitutive Models and Input Motion*

In the FEM phase, the tailings, embankment, and foundation are modeled using the PM4Sand constitutive model, which is designed to characterize potentially liquefiable soils in earthquake applications (Boulanger and Ziotopoulou, 2015). As site-specific measurements of the relative density and shear modulus at the Mochikoshi site are not available, we defer initially to the default PM4Sand parameters, modeling the tailings as a loose sand ($D_R$ = 0.35), the embankment as a medium-dense sand ($D_R$ = 0.55), and the foundation as a dense sand ($D_R$ = 0.75). These parameters were selected based on the general descriptions in Ishihara (1984) and Byrne and Seid-Karbasi (2003) as well as the relative strengths of the different materials. Each material is assigned the unit weight, void ratio, and permeability reported by Okusa and Anma (1979). The PM4Sand constitutive model parameters for the three materials are tabulated in Table 2.

Table 2: Material properties in the FEM phase of the Mochikoshi hybrid model.

| Material | Dr | $G_0$ | $h_{po}$ | Q | R | γ (kN/m³) | e | $k_x$ (m/s) | $k_y$ (m/s) |
|---|---|---|---|---|---|---|---|---|---|
| Tailings | 0.35 | 476 | 2.70 | 11.998 | 3.75 | 18.3 | 0.99 | 7.1E-06 | 7.1E-09 |
| Embankment | 0.55 | 677 | 0.40 | 10 | 1.50 | 16.6 | 1.20 | 1.0E-06 | 1.0E-06 |
| Foundation | 0.75 | 890 | 0.63 | 10 | 1.50 | 18.6 | 0.80 | 1.0E-05 | 1.0E-05 |

Accurately capturing the undrained response of the tailings is critical to modeling the failure process. Therefore, we carefully calibrate the $D_R = 0.35$ parameters for the tailings to ensure that the residual undrained shear strength is consistent with the residual shear strengths of the LEM model (Table 1). Boulanger and Ziotopoulou (2015) calibrated their default PM4Sand parameters to match experimental cyclic resistance curves but acknowledge that they yield unrealistic undrained shear strengths. We follow the calibration procedure outlined by Boulanger and Ziotopoulou (2018) to adjust our model parameters. Our goal is twofold: first, to ensure that the model generates undrained shear strength values that align with our empirical estimates of $s_r$, and second, to maintain cyclic resistance ratio (CRR) curves that approximately match those of the default $D_R = 0.35$ parameters. The calibration is based on simulated undrained cyclic shear tests of a single element, with liquefaction defined as 3% shear strain, and undrained monotonic shear tests to 10% strain to determine the undrained shear strength. The calibration process first adjusts the critical state line (CSL) parameters (Q and R) such that the simulated undrained shear strengths match the target $s_r$ values predicted by Weber (2015) at $\sigma'_{vo}$ of 50, 100, and 250 kPa (Table 3; Figure 5a). These $\sigma'_{vo}$ values are selected because they are representative of the range of confining stresses in the Mochikoshi model. Shifting the CSL changes the CRR curves, so we adjust the contraction parameter ($h_{po}$) until the simulated CRR curves generally match those for the default parameters (Figure 5b). Adjusting $h_{po}$ changes the undrained shear strength, so this process is performed iteratively until both the CRR curve and undrained shear strengths achieve the desired values. Our final calibration results in Q and R values of 11.998 and 3.75, respectively, and $h_{po}$ of 2.7 (Table 2). These parameters are the values at which the undrained monotonic shear strengths (Figure 5a) match the residual strengths in Table 3 and the calibrated CRR curves match those of the default loose sand parameters (Figure 5b).

Table 3: Undrained shear strengths and corresponding consolidation stresses from Weber (2015) to which constitutive model of tailings is calibrated.

| $\sigma'_{vo}$ (kPa) | $s_r$ (kPa) |
|---|---|
| 50 | 4.3 |
| 100 | 6.5 |
| 250 | 11.1 |

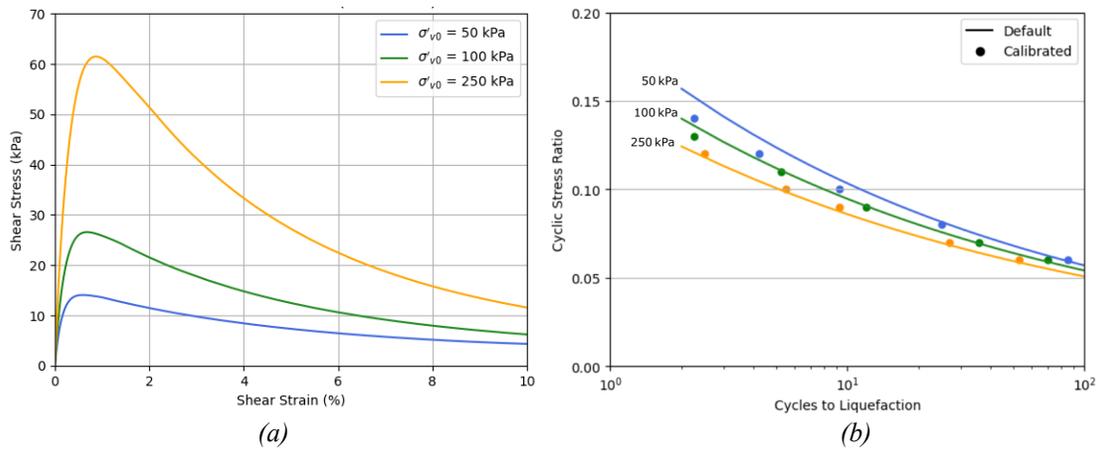

*Figure 5: (a) CRR values of calibrated PM4Sand parameters (points) compared to default parameters (lines) and (b) stress-strain curves of undrained monotonic shear tests of calibrated parameters.*

Due to the absence of recorded time histories from the Izu-Ohshima-Kinkai earthquake, we adopt an approach similar to Byrne and Seid-Karbasi (2003) by using a recording from the 1971 San Fernando earthquake. Specifically, we select the North-South horizontal component recorded at the Caltech Athenaeum Library. To match the estimated peak ground acceleration (PGA) at the Mochikoshi site, which Ishihara (1984) reported to be between 0.2 and 0.3 g, we scale the selected time history to a PGA of 0.3 g. We then convert this acceleration time history (Figure 6a) into a velocity time history (Figure 6b) using numerical integration with the trapezoidal method. This resulting velocity time history is applied to the base of our model to simulate the seismic input.

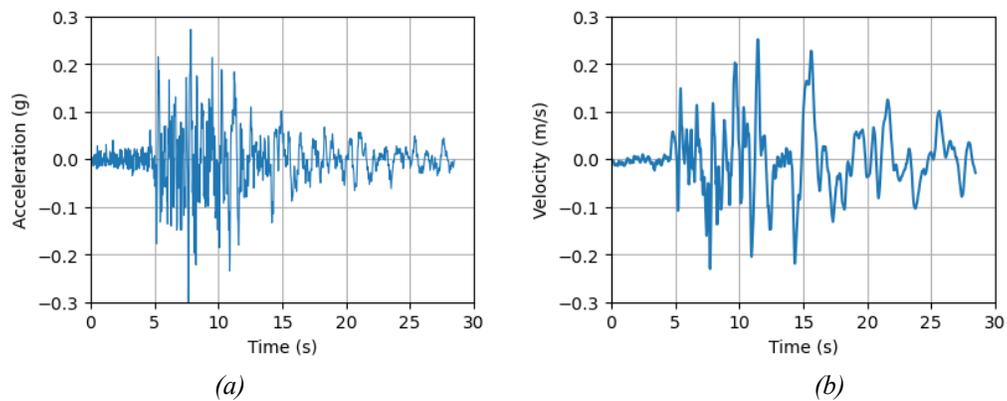

*Figure 6: (a) Acceleration and (b) velocity time history of the input motion applied to the FEM phase of the hybrid model.*

*Failure Initiation in the FEM Phase*

Our analysis begins by examining how the FEM response evolves over time, focusing on three key aspects: movements, pore pressures (i.e., liquefaction), and kinetic energy. To quantify movements, we track the displacement magnitude at the crest of the upper dam. To assess liquefaction and its impact on failure progression, we utilize the excess pore pressure ratio ($r_u$) at each node, defined as:

$$r_u = \frac{\Delta u}{\sigma'_{vo}} \quad (1)$$

where $\Delta u$ is the excess pore pressure and $\sigma'_{vo}$ is the initial effective vertical stress. Under level-ground conditions with cyclic stress reversals, $r_u$ = 1.0 is often associated with liquefaction because it represents a zero effective stress condition. In sloping ground conditions with static shear stress, $r_u$ often does not reach 1.0 due to absence of cyclic stress reversals. Yet, the soil will still experience collapse and lose significant strength even if $r_u$ does not reach 1.0 (Idriss and Boulanger, 2008; National Academies of Sciences, Engineering, and Medicine, 2021). To account for this phenomenon, we consider nodes with $r_u$ > 0.7 as liquefied. We focus our analysis on the area directly beneath the crest of the upper dam, monitoring the maximum depth of liquefied nodes in this zone. This region is particularly critical to the failure surface development, as indicated by our LEM analysis.

Figure 7 illustrates the time evolution of crest displacement, liquefaction depth, and kinetic energy. The dam initially shows only minor displacements, liquefaction depths < 10 m and minimal kinetic energy, but at $t \sim 11$ s the liquefaction depth reaches ~ 12 m, the minimum depth of liquefaction required to achieve FS = 1 in our LEM model. This depth of liquefaction triggers an acceleration in the displacement of the crest and an increase in the kinetic energy, signaling the onset of instability. Although significant shaking ends at $t \sim 13$ s (Figure 6a), displacements keep increasing beyond this time, indicating that the earthquake has initiated a self-propagating failure. At $t \sim 15$ s, we observe rapid acceleration in the kinetic energy and a continued increase in crest displacement (Figure 7). This response corresponds to liquefaction reaching the critical depth of the LEM model (18.5 m), which occurs at $t \sim 14.75$ s in the FEM analysis. This depth of liquefaction yielded a minimum FS of 0.54 and a deeper failure surface in the LEM model. Thus, we interpret the behavior after $t$ > 14.75 s as the mobilization of a larger, deeper failure mass. This mass continues to accelerate until mesh entanglement occurs at $t \sim 23$ s.

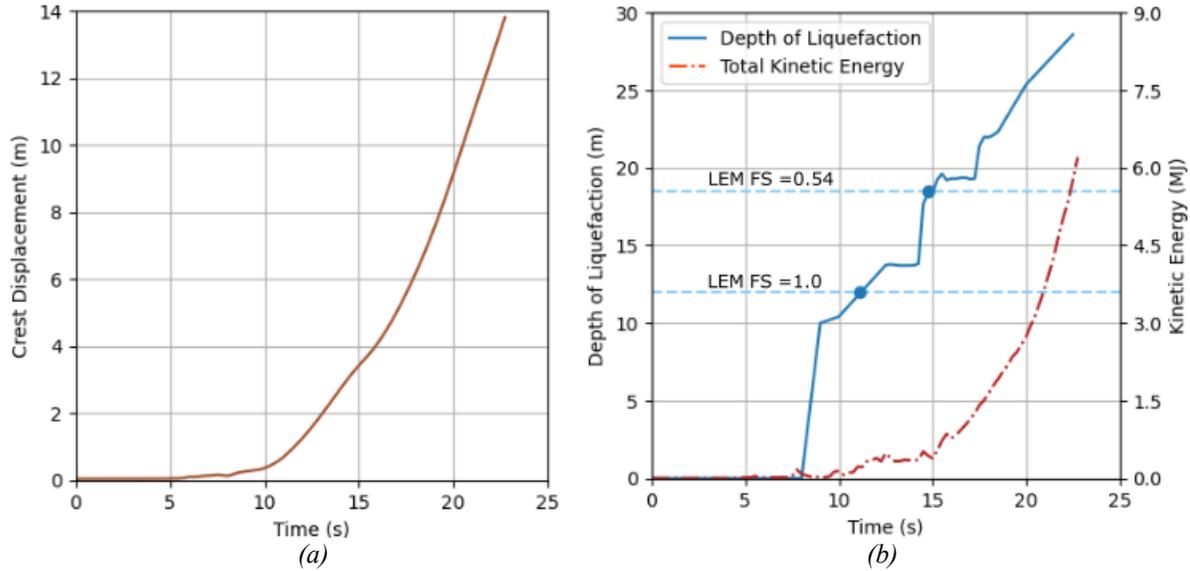

*Figure 7: Time evolution of (a) magnitude of displacement of the crest of the upper dam and (b) depth of liquefaction and total kinetic energy in the FEM phase.*

Figure 8 illustrates the deviatoric shear strain ($\varepsilon_q$) and $r_u$ at key times in the FEM phase: $t = 10$, 13, and 14.75 s. At $t = 10$ s (Figure 8a), we observe an initial shear surface developing, characterized by $\varepsilon_q$ exceeding 10% along a thin zone with $r_u > 0.7$. By $t = 13$ s, this zone of significant deviatoric shearing has expanded to form a full failure surface, accompanied by a corresponding expansion of the $r_u > 0.7$ zone. Notably, a second, deeper failure surface also begins to develop at an elevation of ~10 m. At $t = 14.75$ s, the second failure surface fully develops with $\varepsilon_q > 10\%$ and the liquefaction zone with $r_u > 0.7$ extends to the critical depth of ~18.5 m below the crest. Thus, the rapid increase in kinetic energy observed at $t > 14.75$ s (Figure 7b) coincides with liquefaction extending deeper, creating a second failure surface that mobilizes a larger failure mass.

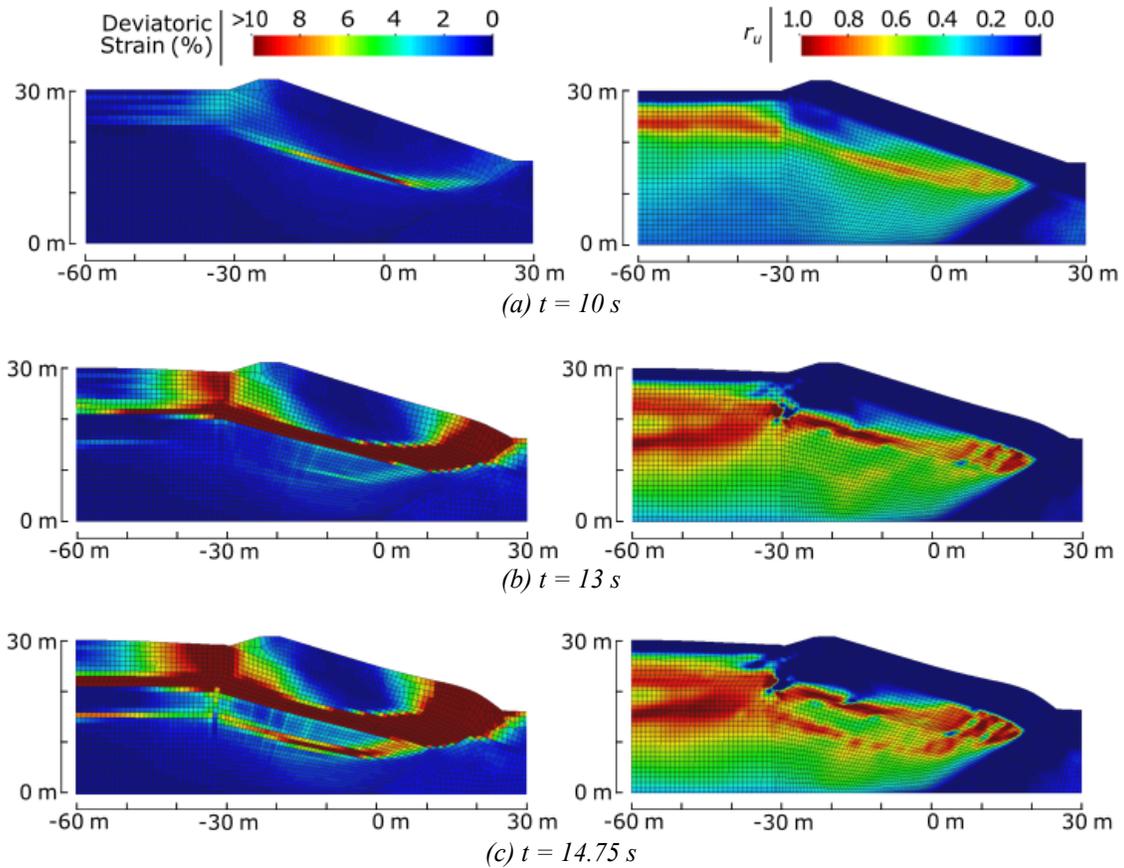

*Figure 8: Contours of deviatoric strain ($\varepsilon_q$; left) and excess pore pressure ($r_u$; right) in the FEM phase at (a) t = 10.0 s and (b) t = 13.0 s and (c) t = 14.75 s, showing the development of the two failure surfaces.*

Analyzing Figures 7 and 8 reveals two critical factors in the failure mechanism: 1) the progression of liquefaction to the critical depth of 18.5 m predicted by our LEM model, and 2) the formation of two distinct failure surfaces. Both these events occur by $t$ = 14.75 s, promptly followed by sharp increases in crest displacement and kinetic energy. This sequence indicates the mobilization of the full failure mass. Consequently, we identify $t$ = 14. 75 s as the point at which the failure mechanism is fully developed, marking the earliest appropriate time for transferring from FEM to MPM analysis.

**TRANSFER FROM FEM TO MPM**

The transfer process involves the transfer of the geometry, material state, and kinematics from the FEM phase to MPM using a Python code that implements the procedure described in Sordo et al. (2024a). Specifically, we convert each FEM element into four MPM material points, initializing them with positions, velocities, stresses, and strains derived from the FEM results. However, our current study presents two challenges not addressed by Sordo et al.

(2024a): (1) how to accurately transfer the liquefaction distribution, and (2) how to select an optimal transfer time for earthquake-induced failures. We discuss these issues in detail in the following sections.

*Transfer of Liquefaction Distribution*

Liquefaction plays a crucial role in the failure of the Mochikoshi dam. We model liquefaction in the MPM phase by identifying a liquefied zone based on the $r_u$ distribution in the FEM model at the transfer time using a threshold of $r_u > 0.7$. This threshold is utilized because cyclic mobility (i.e., the potential for large strains) has been observed in soils in embankment structures at such values of $r_u$ (NASEM, 2014; Rapti et al., 2018). However, the liquefied zone may contain pockets of material with $r_u < 0.7$ that likely will exhibit liquefied behavior during the runout process because they are encapsulated within the liquefied zone with larger $r_u$. These pockets occur primarily below the upper dam (Figure 8c). To capture this behavior in the MPM phase, we consider all tailings particles located above the deepest locations with $r_u > 0.7$ to be liquefied (Figure 9). Unlike the LEM model, where liquefied layers were simplified as horizontal, the liquefied tailings in the MPM phase (Figure 9b) are divided into layers based on the initial vertical stresses from the FEM model, resulting in a more realistic representation of the liquefaction distribution.

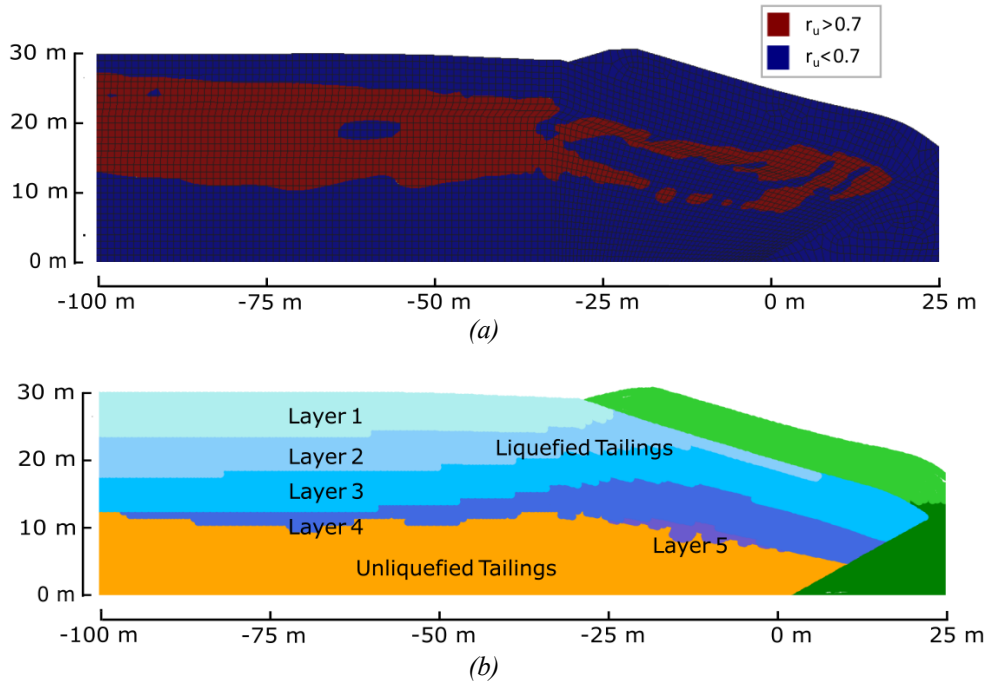

*Figure 9: (a) The $r_u$ > 0.7 fields from the FEM phase at t = 14.75 s (b) the transferred liquefaction distribution for the MPM phase.*

*Selection of Transfer Times*

The ideal $t_T$ should occur after the FEM phase has fully captured the development of the liquefaction-induced failure mechanism, which we preliminarily estimate as $t \geq 14.75$ s. As noted earlier, the MPM phase does not include the seismic loading, so all essential seismic effects must be captured in the FEM phase before transfer. Additionally, prolonged deformation in the FEM phase leads to mesh distortion, which can negatively impact the MPM phase if the transfer occurs too late. Thus, the ideal $t_T$ lies between two conditions: full development of the failure surface and the onset of excessive mesh distortion in the FEM (Figure 1).

Given the lack of strict criteria for identifying the ideal $t_T$ in general cases, we investigate a range of transfer times and evaluate their influence on the runout results. Sordo et al. (2024a) quantified the latest acceptable $t_T$ based on two parameters that represent mesh distortion: the normalized Jacobian determinant (($|J_t| / |J_0|$), where $|J_t|$ and $|J_0|$ are the determinant of an element Jacobian at time $t$ and at $t = 0$ s) and the deviatoric strain ($\varepsilon_q$). These parameters are averaged over elements with $\varepsilon_q > 3\%$ to focus on the failure zone. Smaller values of $(|J_t| / |J_0|)_{avg}$ and larger values of $\varepsilon_{q,avg}$ indicate greater mesh distortion. Sordo et al. (2024a) found that transfers performed before $(|J_t| / |J_0|)_{avg}$ falls below 0.97 and before $\varepsilon_{q,avg}$ reaches 20% yield acceptable errors in the runout results. Figure 10 shows the evolution of these metrics of mesh quality for the Mochikoshi analysis. The $\varepsilon_{q,avg}$ threshold is exceeded at $t \sim 17.5$ s and the $(|J_t| / |J_0|)_{avg}$ threshold is reached at $t \sim 20.25$ s. Ideally, transfers should occur before either of these thresholds is exceeded to minimize mesh distortion effects on the MPM phase.

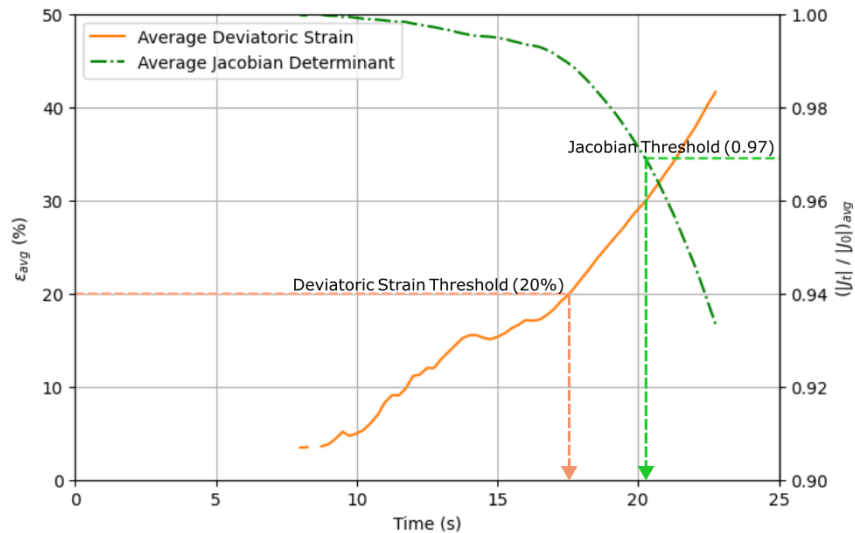

*Figure 10: Average normalized Jacobian determinant (($|J_t| / |J_0|$)$_{avg}$) and average deviatoric strain ($\varepsilon_{q,avg}$) of the FEM phase as a function of time.*

**MPM PHASE OF HYBRID MODEL TO PREDICT RUNOUT**

We perform the runout phase of our hybrid analysis using the CB-Geo MPM code (Kumar et al., 2019). The MPM phase employs a structured background grid composed of square 2.5 m x 2.5 m GIMP cells (Bardenhagen and Kober, 2004) and a timestep of 5e-5 s. This cell size is large enough to ensure that there are at least four particles per cell on the left edge of the model, where the particles are most sparse, a minimum concentration recommended by Soundararajan (2015). Furthermore, GIMP uses higher-order shape functions for stress interpolation to the material points to help reduce stress checkerboarding and minimize cell crossing noise, common issues in MPM simulations.

Our model setup, shown in Figure 11, includes a roller boundary on the left edge of the grid, with the mesh extending 2,000 m to the right to accommodate large runout distances. While the actual Mochikoshi dam failure occurred on a mountainside, resulting in downslope debris flow, we use a flat downstream boundary in our MPM model. This simplification is necessary because GIMP is only compatible with square cells, and our attempts to use non-square, isoparametric cells to model the inclined topography led to numerical instabilities.

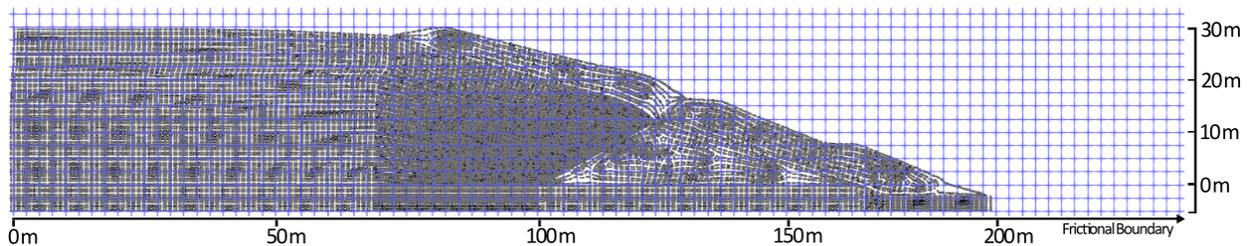

*Figure 11: Computational grid and initial particle positions of the MPM phase immediately after $t_T = 14.75$ s transfer.*

The bottom boundary of our model incorporates frictional properties. We assign a friction coefficient of 0.7 to the area beneath the embankment and tailings pond. For the region downstream of the dam, we use a reduced friction coefficient of 0.35. This lower value accounts for the observed behavior of saturated debris flows, which often experience reduced basal friction due to shear-induced excess pore pressure generation (Cleary and Campbell, 1993; Breard et al., 2020).

The modeling choices described above allow us to simulate the runout phase while balancing computational stability and real-world behavior representation. Although our flat downstream boundary is a simplification of the actual topography, it enables us to capture the essential dynamics of the failure and runout process.

*Material Properties for the MPM Phase*

In the MPM phase, we simplify our constitutive modeling approach compared to the FEM phase. We opt for a total stress Mohr-Coulomb model for the tailings and dam materials, rather than the more complex PM4Sand model used in the FEM phase. This simplification is justified because excess pore pressure development is no longer relevant after the strong shaking has ended, and MPM is inherently prone to stress inaccuracies. This approach, while a simplification, has been used by other researchers for MPM runout modeling (e.g., Macedo et al., 2024; Talbot et al., 2024). For the foundation soil, which does not participate in the runout process, we use a linear-elastic model. We maintain consistency with the FEM phase by using the same unit weights for all materials (Table 2).

Table 4 summarizes the material properties used in the MPM phase. The initial Mohr-Coulomb strength parameters ($c_{init}$, $\phi_{init}$) are the peak strengths for the embankment and unliquefied tailings (Table 1), and $c_{init} = s_r$ with $\phi_{init} = 0$ for the different layers of liquefied tailings. Thus, the liquefied tailings are effectively analyzed via a total stress undrained constitutive model. Only zones of the tailings that are identified as liquefied in the FEM are assigned $s_r$. Additional softening of the strength parameters during runout for the liquefied tailings and embankment is considered.

*Table 4: Material properties in the MPM phase of the hybrid model.*

| Material | $\nu$ | E (MPa) | $\phi_{init}$ (°) | $c_{init}$ (kPa) | $\varepsilon_{q,init}$ (%) | $\varepsilon_{q,soft}$ (%) | $\phi_{soft}$ (°) | $c_{soft}$ (kPa) |
|---|---|---|---|---|---|---|---|---|
| Embankment | | | | | | | | |
|     Saturated | 0.492 | 142 | 35 | 25 | 1 | 5 | 20 | 1 |
|     Unsaturated | 0.300 | 124 | 35 | 25 | 1 | 5 | 20 | 1 |
| Unliquefied Tailings | 0.495 | 120 | 30 | 1 | n/a | n/a | n/a | n/a |
| Liquefied Tailings | | | | | | | | |
|     Layer 1 | 0.495 | 72 | 0 | 3 | 0 | 100 | 0 | 1.5 |
|     Layer 2 | 0.495 | 124 | 0 | 5.6 | 0 | 100 | 0 | 2.8 |
|     Layer 3 | 0.495 | 160 | 0 | 7.4 | 0 | 100 | 0 | 3.7 |
|     Layer 4 | 0.495 | 189 | 0 | 9.1 | 0 | 100 | 0 | 4.55 |
|     Layer 5 | 0.495 | 215 | 0 | 10.6 | 0 | 100 | 0 | 5.3 |
| Foundation | 0.491 | 316 | | | linear elastic | | | |

The large-strain behavior of liquefied tailings during runout is not well understood. Factors such as void ratio, fines content, and deposition method significantly influence the shear strength and may not remain constant (Yamamuro and Covert, 2001; Sitharam et al., 2013; Riveros and Sadrekarimi, 2021). To account for potential strength loss due to mechanisms such as void redistribution and water film development during runout (Kokusho and Kojima, 2002), we implement a softened strength represented by a 50% reduction in residual strength, setting $c_{soft} = 0.5 \cdot s_r$ (Table 4). This strength reduction is in addition to the strength loss due to pore pressure generation during liquefaction.

We also consider softening in the embankment materials. Ishihara (1984) described observations of fissure formation in the Mochikoshi embankments during the earthquake, allowing pore water from the tailings to infiltrate previously dry sections. This suggests possible fluidization of the embankment material during breach, a process where rapidly sheared coarse, contractive soils lose significant strength as shearing outpaces pore water dissipation (Okura et al., 2002). To model this, we assign softened strength properties of $c_{soft} = 1$ kPa and $\phi_{soft} = 20°$, representing complete loss of cohesion and a 50% reduction in frictional strength. Without this softening of the embankment materials, the tailings material did not experience large-scale runout.

The softening of $c$ and $\tan\phi$ occurs linearly from their initial values ($c_{init}$ and $\tan\phi_{init}$) to their softened values ($c_{soft}$ and $\tan\phi_{soft}$) as a function of the deviatoric strain ($\varepsilon_q$). The softening begins at $\varepsilon_{q,init}$ and finishes at $\varepsilon_{q,soft}$. For the liquefied tailings the values of $\varepsilon_{q,init}$ and $\varepsilon_{q,soft}$ are 0% and 100%, respectively, and for the embankment materials they are 1% and 5% (Table 4).

*Runout Results of the MPM Phase*

Figure 12 shows the final MPM runout geometries for transfer times $t_T = 13$ s and $t_T = 16$ s, both incorporating strength softening. The earlier transfer time, $t_T = 13$ s, results in ~100 m of runout, while the later transfer at $t_T = 16$ s produces ~200 m of runout. This difference arises because the $t_T = 13$ s model is transferred before the full development of the failure mechanism (Figure 8b), In contrast, the $t_T = 16.0$ s model is transferred after complete failure mechanism development (Figure 8c) but before excessive mesh distortion occurs (as indicated by the metrics in Figure 10), resulting in much larger runout. These results clearly demonstrate the significant impact of transfer time on computed runout distances.

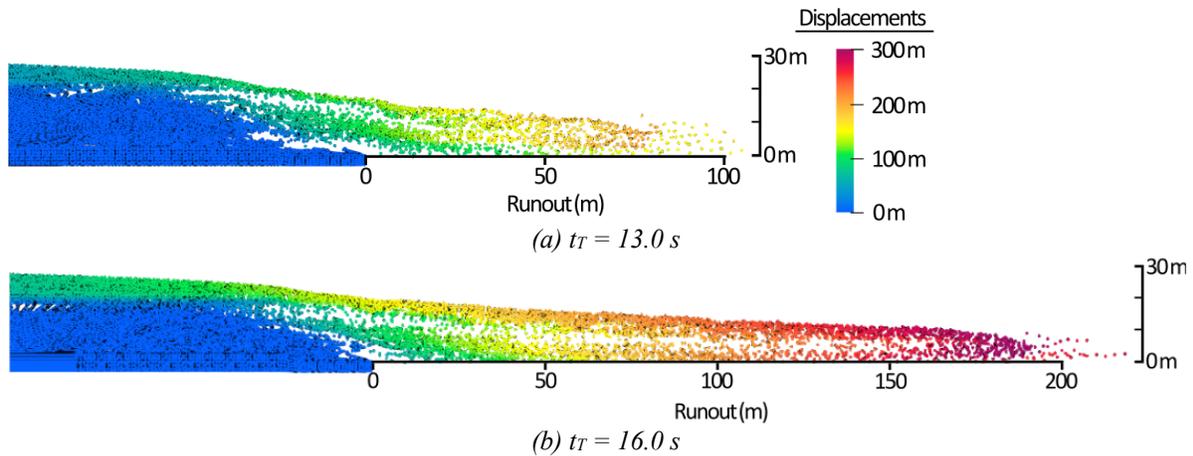

*Figure 12: Example of the final runout geometries from the (a) $t_T$ = 13.0 s and (b) $t_T$ = 16.0 s hybrid models.*

To thoroughly investigate the influence of transfer time on runout, we conducted MPM analyses for $t_T$ ranging from 10 and 22.5 s. These analyses were performed both with and without softening of the residual strength of the liquefied tailings. All analyses included softening of the embankment material, as runout did not occur without this feature. To prevent overestimation of runout due to a few unstable particles, we define the runout distance as the 99th percentile of particle distances from the toe of the lower dam.

Figure 13 presents the relationship between final runout distance and transfer time, and results are shown for analyses with and without residual strength softening of the liquefied tailings. Transfers before $t_T$ = 12.5 s produced no runout, while transfers between $t_T$ = 12.75 and 14.25 s yielded some runout but only when residual strength softening was included in the liquefied tailings. Runout significantly increases when $t_T \geq 14.5$ s. In cases without softening, these are the first transfers that yield any runout. Transfers between 14.5 and 15.0 s appear to constitute a transition period where runout significantly increases. For $t_T \geq 15.25$ s, the models with and without softening both yield their maximum runout distances. These runout distances are ~35 m for the models without residual strength softening and ~215 m for the models with residual strength softening (Figure 13). These maximum runout values remain consistent for transfers between $t_T \sim 15.25$ and 20 s, after which runout distances decrease. The final transfer ($t_T$ = 22.5 s), performed just before the mesh entangles, resulted in a 25% reduction in runout distance for the softened model and a 40% reduction for the model without softening.

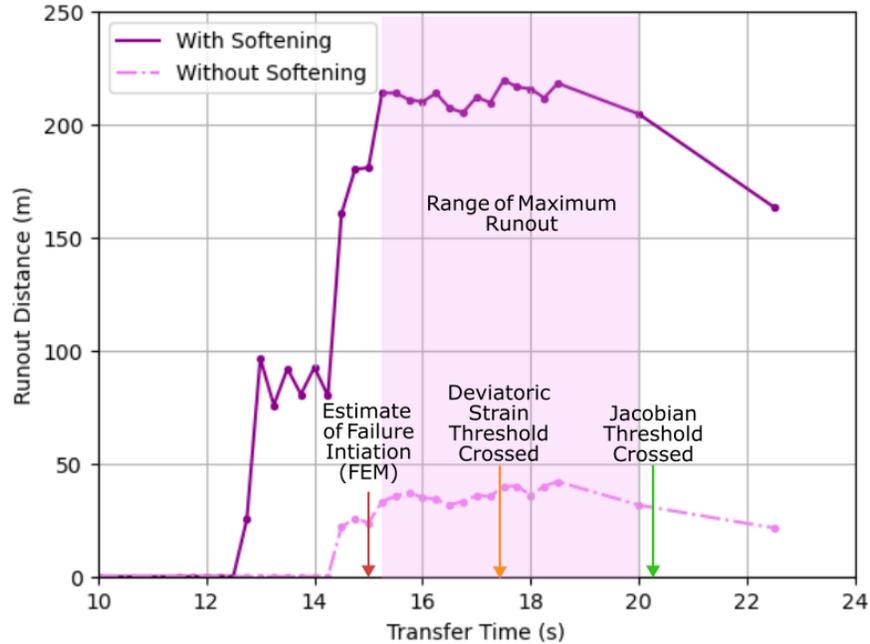

*Figure 13: Runout results as a function of transfer time and softening state of the liquefied tailings.*

Based on the FEM phase conditions, we initially estimated that the failure mechanism would be fully developed, and thus maximum runout achieved, when $t_T \geq 14.75$ s. Transfers before this time indeed produced reduced or no runout due to incomplete development of the failure mechanism. However, the maximum runout was not achieved until $t_T \geq 15.25$, indicating our initial estimate was slightly early. We therefore recommend selecting a transfer time somewhat later than the earliest estimate of the development of the full failure mechanism.

The observed decrease in runout for later transfers aligns with findings from Sordo et al. (2024a), where mesh distortion progressively reduces runout. Figure 14 shows the FEM mesh deformations at various times approaching mesh entanglement. The deformed shape at $t = 17.5$ s (Figure 14a) appears realistic and in line with the case history description of bulging of the upper dam prior to breach (Ishihara, 1984). However, by $t = 20.0$ s (Figure 14b) the bulging of the lower dam becomes unrealistic, and by $t = 22.5$ s (Figure 14c) it is significantly so. These unrealistic deformations stem from limitations in handling large deformations in FEM, similar to the extreme toe bulging observed in hybrid slope failure models from Sordo et al. (2024a). Transfers after $t = 20.0$ s produce deteriorating runout results due to these unrealistic deformations. Interestingly, this deterioration begins later than the threshold (i.e., $\varepsilon_{q,\text{avg}} > 20\%$ or $(|J_t| / |J_0|)_{\text{avg}} < 0.97$) predicted by Sordo et al. (2024a). Our model reaches these criteria at $t = 17.5$ and $20.25$ s, respectively (Figure 10), but Sordo et al. (2024a) predicted that the deterioration of runout results would begin after either of these criteria are exceeded. We only observe decreased runout for $t_T > 20.0$ s suggesting that the Jacobian threshold may be a better indicator of the latest acceptable transfer time in this application.

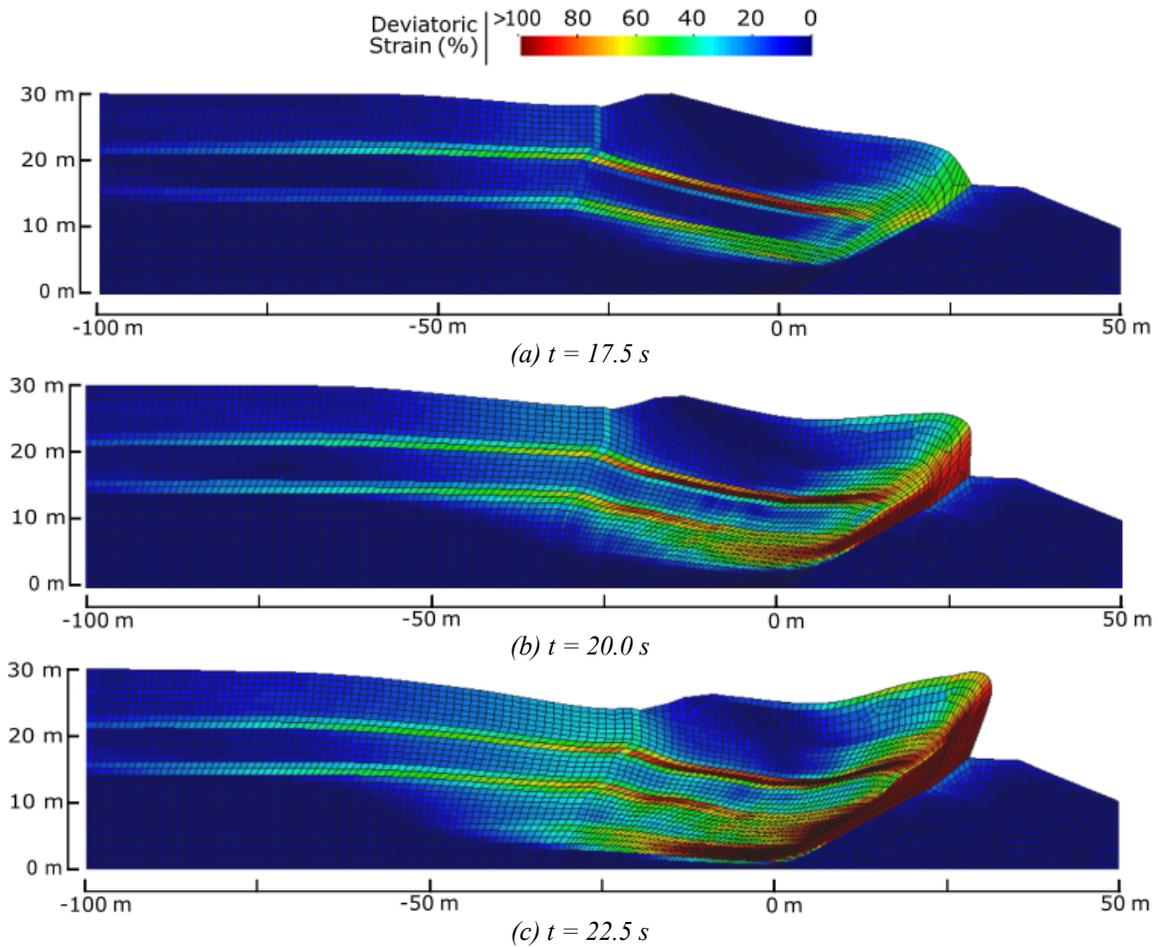

Figure 14: Deviatoric strain contours of the deformed FEM mesh at (a) 17.5 s, (b) 20.0 s, and (c) 22.5 s.

Finally, the inclusion of strain softening significantly increases maximum runout distances. Models with 50% softening in residual strength produce maximum runout distances 5 to 6 times greater than those without softening. While these results highlight the substantial influence of the strength of the liquefied tailings on runout distance, they also show that the ideal transfer time remains consistent regardless of softening. This underscores the importance of understanding material behavior in both the initiation (small-strain) and runout (large-strain) phases of failure, and the need for constitutive models that accurately capture this behavior. For more accurate runout analyses, further research is needed to better characterize the large-strain strength of liquefied materials during runout.

**CONCLUSIONS**

This study demonstrates the unique capabilities of a sequential hybrid finite element method (FEM) and material point method (MPM) approach to effectively simulate tailings dam failures with complex triggering mechanisms from initiation to runout. FEM excels at predicting the seismic response of a site, including the triggering of liquefaction, but is limited in modeling large deformations because of the effects of mesh distortion. Conversely, MPM can handle large deformations but lacks the accuracy of FEM for simulating complex failure mechanisms. By coupling these two methods sequentially, the hybrid FEM-MPM approach leverages the strengths of each numerical technique to simulate the distinct stages of failure and can thus predict the conditions that will initiate a tailings dam failure and offer an estimate of the consequences. To demonstrate this hybrid FEM-MPM method, we apply it to the 1978 Mochikoshi tailings dam failure in Japan, which failed due to the combined effects of seismic loading and liquefaction.

The FEM analysis of the Mochikoshi failure involved the characterization of the liquefaction response of the tailings, both in terms of the occurrence of liquefaction and the post-liquefaction residual strength. For this purpose, the PM4Sand model was calibrated to achieve appropriate cyclic resistance curves and residual strengths for a loose sand, as predicted by empirical relationships. An earthquake motion scaled to PGA = 0.3 g was used to trigger failure in the FEM analysis. A separate LEM model was analyzed to understand the relationship between the depth of liquefaction and instability of the tailings dam. These analyses showed that a factor of safety of 1.0 was achieved with liquefaction extending about 12 m deep, and a minimum factor of safety of 0.54 occurred when liquefaction extended about 18.5 m. The movements and kinetic energy of the tailings dam accelerated when the liquefaction in the FEM analysis reached the critical depth of 18.5 m, and two shear surfaces emerged. The zone of liquefaction was identified from the FEM analysis and transferred to the MPM phase based on $r_u > 0.7$, with this zone assigned stress-dependent, liquefied residual strengths. The MPM phase received initial state variables based on the FEM phase and was then simulated under static conditions (i.e., no earthquake shaking) with strengths assigned as Mohr-Coulomb parameters to predict the final runout.

The transition from FEM to MPM is a critical step requiring careful timing of the transfer. An ideal transfer window exists between the full development of the failure initiation in FEM and the onset of excessive mesh distortion. If the transfer is performed before this ideal transfer window, the failure mechanism is not fully captured in FEM. Conversely, transferring after this ideal window results in inheriting the results of mesh distortion from FEM. Both of these errors impair the accuracy of the final results by underestimating runout. The limits of this ideal transfer window can be identified based on the conditions of the FEM phase. For the analyses performed in this study, the full development of the liquefaction-induced failure was indicated by the formation of distinct failure surfaces, liquefaction extending to a critical depth of ~18.5 m, and a significant increase in kinetic energy signaling the onset of progressive failure. These criteria were met at $t \sim 14.75$ s. Mesh distortion accumulates gradually in the FEM analysis, and it was measured using the average normalized Jacobian determinant, $(|J_t| / |J_0|)_{avg}$, and the average deviatoric strain, $\varepsilon_{q,avg}$. These parameters were averaged only over elements with $\varepsilon_q > 3\%$. Although previous work indicated both $\varepsilon_{q,avg}$ and $(|J_t| / |J_0|)_{avg}$ affect runout, the results of this study showed that runout was affected only when $(|J_t| / |J_0|)_{avg}$ fell below 0.97. This criterion was met at $t \sim 20.25$ s. Runout results for analyses performed for a range of transfer times indicated similar maximum runouts for transfers between $t = 15.25$ and 20 s. The lower limit of this ideal transfer time window is slightly later than the estimate of ~14.75 s, and the upper limit of this window corresponds well to the $(|J_t| / |J_0|)_{avg}$ threshold. It is recommended that, in application of the hybrid method, a transfer time be selected approximately halfway between the lower and upper limits.

The computed runout is significantly affected by the characterization of the softening response of the liquefied tailings and embankment. Runout was experienced only when the embankment materials were allowed to soften. Furthermore, softening of the residual strength of the liquefied tailings by 50% increased the runout more than five-fold, from about 35 m to 215 m. Only when this softening was included was the predicted runout within the order of magnitude of the case history. However, softening of the liquefied strengths in the runout phase does not appear to affect the ideal window of transfer times.

Overall, this work highlights the potential of hybrid numerical techniques to overcome the limitations of individual methods and provide comprehensive simulations of complex geotechnical failures. The proposed FEM-MPM approach offers a robust tool for analyzing complex tailings dam failures to inform risk assessment and mitigation strategies.

**DATA AVAILABILITY STATEMENT**

The FEM and MPM models of the Mochikoshi tailings dam, the scripts used to perform the FEM to MPM transfer and analyze data, and all simulation results from this work are available electronically at Sordo et al. (2024b).

# REFERENCES

Adamo, N., Al-Ansari, N., Sissakian, V., Laue, J., & Knutsson, S. (2020). Dam Safety: The Question of Tailings Dams. *Journal of Earth Sciences and Geotechnical Engineering*, *11*(1), 1–26.

Acosta, J., Vardon, P., Remmerswaal, G., & Hicks, M. (2020). An Investigation of Stress Inaccuracies and Proposed Solution in the Material Point Method. *Computational Mechanics*, 65, 555-581.

Alsardi, A., & Yerro, A. (2023). Coseismic Site Response and Slope Instability Using Periodic Boundary Conditions in the Material Point Method. *Journal of Rock Mechanics and Geotechnical Engineering*, *15*(3), 641–658.

Bardenhagen, S., Brackbill, J., & Sulsky, D. (2000). The Material-Point Method for Granular Materials. *Computer Methods in Applied Mechanics and Engineering*, *187*(3–4), 529–541.

Bardenhagen, S., & Kober, E. (2004). The Generalized Interpolation Material Point Method. *Tech Science Press*, *5*(6), 477–495.

Bentley Systems Incorporated. *Plaxis LE Slope Stability*. (2021). [Computer software]. https://communities.bentley.com/cfs-file/__key/communityserver-wikis-components-files/00-00-00-05-58/8585.SlopeStability_5F00_Theory_5F00_Manual.pdf

Boulanger, R., & Ziotopoulou, K. (2015). *PM4SAND (Version 3): A Sand Plasticity Model for Earthquake Engineering Applications*. UC Davis Center for Geotechnical Modeling.

Boulanger, R., & Ziotopoulou, K. (2018). *On NDA Practices for Evaluating Liquefaction Effects* (Geotechnical Special Publication 290). ASCE; Proc., Geotechnical Earthquake Engineering and Soil Dynamics V.

Breard, E., Dufek, J., Fullard, L., & Carrara, A. (2020). The Basal Friction Coefficient of Granular Flows With and Without Excess Pore Pressure: Implications for Pyroclastic Density Currents, Water-Rich Debris Flows, and Rock and Submarine Avalanches. *Journal of Geophysical Research*, *125*.

Byrne, P., & Seid-Karbasi, M. (2003). "Seismic Stability of Impoundments." 17th Annual Symposium, Vancouver Geotechnical Society, Vancouver, Canada.

Cleary, P., & Campbell, C. (1993). Self-Lubrication of Long Runout Landslides: Examination by Computer Simulation. *Journal of Geophysical Research*, *98*(B12), 21911–21924.

Cuomo, S., Di Perna, A., & Martinelli, M. (2021). Modeling the Spatio-Temporal Evolution of Rainfall-Induced Retrogressive Landslide in an Unsaturated Slope. *Engineering Geology*, *294*(3).

Fourie, A., Verdugo, R., Bjelkevik, A., Torres-Cruz, L., & Znidarcic, D. (2022). *Geotechnics of Mine Tailings: A 2022 State of the Art*. 20th International Conference on Soil Mechanics and Geotechnical Engineering, Sydney, Australia.

Geppetti, A., Facciorusso, J., & Madiai, C. (2022). *Tailings Dams Numerical Models: A Review*. 7th World Congress on Civil, Structural, and Environmental Engineering, Virtual Conference.

Griffiths, D., & Lane, P. (1999). Slope Stability Analysis by Finite Elements. *Geotechnique*, *49*(3), 387–403.

Hoang, T., Nguyen, T. T., Nguyen, T. V., Nguyen, G., Bui, H. (2024). SPH Simulation of Earthquakr-Induced Liquefaction and Large Deformation Behavior of Granular Materials Using SANISAND Constitutive Model. *Computers and Geotechnics*, 174.

Hoang, T., Bui, H., Nguyen, T. T., Nguyen, T. V., Nguyen, G. (2024). Development of Free-Field and Compliant Base SPH Boundary Conditions for Large Deformation Seismic Response Analysis of Geomechanics Problems. *Computer Methods in Applied Mechanics and Engineering, 432(A)*.

ICMM. (2020). *Global Industry Standard on Tailings Management*. International Council on Mining and Metals.

Idriss, I., & Boulanger, R. (2008). *Soil Liquefaction During Earthquakes*. Oakland, CA: Earthquake Engineering Research Institute.

Ishihara, K. (1984). *Post-Earthquake Failure of a Tailings Dam Due to Liquefaction of Pond Deposit. 13*.

Kohler, M., Stoecklin, A., Puzrin, A. (2021). A MPM Framework for Large-Deformation Seismic Response Analysis. *Candian Geotechnical Journal, 59(6)*.

Kokusho, T., & T. Kojima. (2002). Mechanism for Postliquefaction Water Film Generation in Layered Sand. *Journal of Geotechnical and Geoenvironmental Engineering, 128(2)*.

Kossoff, D., Dubbin, W., Alfredsson, M., Edwards, S., Macklin, M., & Hudson-Edwards, K. (2014). Mine Tailings Dams: Characteristics, Failure, Environmental Impacts, and Remediation. *Applied Geochemistry*, *51*, 229–245.

Kramer, S., & Wang, C. (2015). Empirical Model for Estimation of the Residual Strength of Liquefied Soil. *Journal of Geotechnical and Geoenvironmental Engineering*, *141*(9).

Kumar, K., Salmond, J., Kularathna, S., Wilkes, C., Tjung, E., Biscontin, G., & Soga, K. (2019). Scalable and modular material point method for large scale simulations. 2nd International Conference on the Material Point Method. Cambridge, UK. https://arxiv.org/abs/1909.13380

Kurima, J., Chandra, B., Soga, K. (2024). Absorbing Boundary Conditions in Material Point Method Adopting Perfectly Matched Layer Theory. 10.48550/arXiv.2407.02790

Larrauri, P., & Lall, U. (2018). Tailings Dams Failures: Updated Statistical Model for Discharge Volume and Runout. *Environments*, *5*(2).

Liang, W., & Zhao, J. (2018). Multiscale Modeling of Large Deformation in Geomechanics. *International Journal for Numerical and Analytical Methods in Geomechanics*, *43*(5), 1080–1114.

Lu, M., Ceccato, F., Zhou, M., Yerro, A., Zhang, J. (2023). Evaluating the Exceedance Probability of the Runout Distance of Rainfall-Induced Landslides Using a Two-Stage FEM-MPM Approach. *Acta Geotechnica*, *19*, 3691-3706.

Lysmer, J., & Kuhlemeyer, R. (1969). Finite Dynamic Model for Infinite Media. *Journal of the Engineering-Mechanics Division*, *95*(4).

Lyu, Z., Chai, J., Xu, Z., Qin, Y., & Cao, J. (2019). A Comprehensive Review on Reasons for Tailings Dam Failures Based on Case History. *Advances in Civil Engineering*, *2019*.

Macedo, J., Yerro, A., Cornejo, R., & Pierce, I. (2024). Cadia TSF Failure Assessment Considering Triggering and Posttriggering Mechanisms. *Journal of Geotechnical and Geoenvironmental Engineering*, *150*(4).

Martin, V., Fontaine, D., & Cathcart, J. (2015). *Challenges with Conducting Tailings Dam Breach Studies*. Tailings and Mine Waste, Vancouver, BC.

McGann, C., Arduino, P., & Mackenzie-Helnwein, P. (2012). Stabilized Single-Point 4-Node Quadrilateral Element for Dynamic Analysis of Fluid Saturated Porous Media. *Acta Geotechnica*, *7*(4), 297–311.

McKenna, F. (1997). *Object-Oriented Finite Element Programming: Frame-Works for Analysis, Algorithms and Parallel Computing* [PhD Dissertation]. UC Berkeley.

National Academies of Sciences, Engineering, and Medicine. 2021. *State of the Art and Practice in the Assessment of Earthquake-Induced Soil Liquefaction and Its Consequences*. Washington, DC: The National Academies Press. https://doi.org/10.17226/23474.

Okura, Y., Kitahara, H., Ochiai, H., Sammori, T., & Kawanami, A. (2002). Landslide Fluidization Process by Flume Experiments. *Engineering Geology*, *66*, 65–78.

Okusa, S., & Anma, S. (1979). Slope Failures and Tailings Dam Damage in the 1978 Izu-Ohshima-Kinkai Earthquake. *Engineering Geology*, *16*, 195–224.

28 of 30

Rapti, I., Lopez-Caballero, F., Modaressi-Farahmand-Razavi, A., Foucault, A., Voldoire, F. (2018). Liquefaction Analysis and Damage Evaluation of Embankment-Type Structures. *Acta Geotechnica*, *13*, 1041-1059.

Ribó, R., de Riera Paseanu, M., & Escolano, E. (1999). *GiD User Manual*. International Center for Numerical Methods and Engineering.

Riveros, G., & Sadrekarimi, A. (2021). Static Liquefaction Behavior of Gold Mine Tailings. *Canadian Geotechnical Journal*, *58*(6).

Sadeghirad A., Brannon R.M., & Burghardt J. (2011). A Convected Particle Domain Interpolation Technique to Extend Applicability of the Material Point Method for Problems Involving Massive Deformations. *International Journal for Numerical Methods in Engineering, 86(12),* 1435-1456.

Sarantonis, E., Etezad, M., & Ghafghazi, M. (2020). The Effect of Assumed Residual Strength on Remediation Cost of a Typical Tailings Dam. *Proceedings Tailings and Mine Waste 2020*. Tailings and Mine Waste, Fort Collins, Colorado.

Seed, H., Idriss, I., Lee, K., & Makdisi, F. (1975). Dynamic Analysis of the Slide in the Lower San Fernando Dam during the Earthquake fo February 9, 1971. *Journal of the Geotechnical Engineering Division*, *101*(9).

Shan, Z., Z. Liao, Y. Dong, D. Wang, & L. Cui. (2021). Implementation of Absorbing Boundary Conditions in Dynamic Simulation of the Material Point Method. *Journal of Zhejiang University - Science A (Applied Physics & Engineering).* 22(11), 870-881.

Sitharam, G., Dash, H., & Jakka, R. (2013). Postliquefaction Undrained Shear Behavior of Sand-Silt Mixtures at Constant Void Ratio. *International Journal of Geomechanics*, *13*(4).

Sordo, B., Rathje, E., & Kumar, K. (2024). Sequential Hybrid Finite Element and Material Point Method to Simulate Slope Failures. *Computers and Geotechnics, 173*.

Sordo, B., E. Rathje, K. Kumar (2024). "Mochikoshi Tailings Dam Simulation", in *Analysis of Mochikoshi Dam Failure using a Hybrid FEM-MPM Approach*. DesignSafe-CI. https://doi.org/10.17603/ds2-32fp-8535

Soga, K., Alonso, E., Yerro, A., Kumar, K., & Bandara, S. (2015). Trends in Large-Deformation Analysis of Landslide Mass Movements with Particular Emphasis on the Material Point Method. *Geotechnique*, *66*(3), 248–273.

Soundararajan, K. K. (2015). Multi-Scale Multiphase Modelling of Granular Flows (Doctoral Thesis). University of Cambridge. https://doi.org/10.17863/CAM.14130.

Steffen, M., Kirby, R., Berzins, M. (2008). Analysis and Reduction of Quadrature Errors in the Material Point Method (MPM). *International Journal of Numerical Methods in Engineering, 76(6)*, 922-948.

Sulsky, D., Chen, Z., & Schreyer, H. (1994). A Particle Method for History Dependent Material. *Computer Methods in Applied Mechanics and Engineering*, *118*(1–2), 179–196.

Talbot, L., Given, J., Tjung, E., Liang, Y., Chowdhury, K., Seed, R., & Soga, K. (2024). Modeling Large-Deformation Features of the Lower San Fernando Dam Failure with the Material Point Method. *Computers and Geotechnics*, *165*.

Wang, L., Coombs, W., Augarde, C., Cortis, M., Brown, M., Brennan, A., Knappett, J., Davidson, C., Richards, D., White, D., & Blake, A. (2021). An Efficient and Locking-Free Material Point Method for Three-Dimensional Analysis with Simplex Elements. *International Journal for Numerical and Analytical Methods in Geomechanics*, *122*, 3876–3899.

Weber, J. (2015). *Engineering Evaluation of Post-Liquefaction Strength* [PhD Dissertation]. UC Berkeley.

Yamamuro, J., & Covert, K. (2001). Monotonic and Cyclic Liquefaction of Very Loose Sands with High Silt Content. *Journal of Geotechnical and Geoenvironmental Engineering*, *127*(4), 314–324.

Youd, T., Idriss, I., Andrus, R., Arango, I., Castro, G., Christian, J., Dobry, R., Finn, W., Harder, L., Hynes, M., Ishihara, K., Koester, J., Liao, S., Marcuson, W., Martin, G., Mitchell, J., Martin, G., Mitchell, J., Moriwaki, Y., Power, M., Robertson, P., Seed, R., Stokoe, K. (2001). Liquefaction Resistance of Soils: Summary Report from the 1996 NCEER and 1998 NCEER/NSF Workshops on Evaluation of Liquefaction Resistance of Soils. *Journal of Geotechnical and Geoenvironmental Engineering*, *127*(10), 817-833.

Zeng, W., & Liu, G. (2018). Smoothed Finite Element Methods (S-FEM): An Overview and Recent Developments. *Archives of Computational Methods in Engineering*, *25*, 397–435.

Zhang, X., Chen, Z., & Liu, Y. (2016). *The Material Point Method: A Continuum-Based Particle Method for Extreme Loading Cases*. Academic Press.

Zienkiewicz, O., Taylor, R., & Zhu, J. (2005). *The Finite Element Method: Its Basis and Fundamentals* (6th ed.). Elsevier Butterworth-Heinemann.